\documentclass[preprint]{elsarticle}
\usepackage{amssymb}
\usepackage{amsthm}

\usepackage{lineno}

\journal{Physica E}

\begin{document}

\begin{frontmatter}

\title{Light Induced Negative Differential Conductance in Molecular Junctions: Role of Triplet States and Electron-Phonon Interaction}

\author[iams]{Amir Eskandari-asl\corref{cor1}}
\ead{amir.eskandari.asl@gmail.com}
\address[iams]{Institute of Atomic and Molecular Sciences, Academia Sinica, Taipei 10617, Taiwan}

\begin{abstract}
In this work we theoretically consider the OPE-3 molecule bridging two metallic leads and show that because of the electron-phonon interaction and the transition of cation to triplet states, we can have a light induced negative differential conductance. Furthermore, we investigate the effects of light intensity and temperature on this phenomena.
\end{abstract}

\begin{keyword}
negative differential conductance \sep triplet states \sep electron-phonon interaction \sep light induced transition \sep master equation
\end{keyword}

\end{frontmatter}


\section{introduction}
In the last two decades advances in technology made the realization of molecular junctions and measuring the charge transport through them possible \cite{strachan2005, cui2001, park2002, xu2003, kubatkin2003, dadosh2005, venka2006, moth2009, xin2019, morteza2018}. At molecular scales the interaction of electrons with the quantum of the vibrational modes, the so called phonons, plays an important role in determining the transport properties. Several novel phenomena including negative differential conductance (NDC), dynamical switching and current hysteresis\cite{xu2015negative,CLi1,Elor,Lilj,le2003negative}, are believed to originate from electron-phonon (e-ph) interaction\cite{galperin2005,galperin2008non,eskandari2016bi1,eskandari2019interplay}.
 
One interesting issue which attracted a huge interest recently, is optical control of the electron transport through molecular junctions\cite{yoshida2015terahertz,jia2016covalently,zhou2018photoconductance}. Of special interest is to have a light induced current, in which the current can be switched on and off upon illumination. This can be achieved either by exciting the electrons in the case that the incoming light frequency is in resonance with some energy gaps of the molecular states\cite{zhou2018photoconductance,petrov2013transient,kornbluth2013light,fainberg2012photoinduced,battacharyya2011optical}, or by using a high intensity coherent quantum light to induce photon side bands that open new transport channels, the so called photon-assisted tunneling\cite{yoshida2015terahertz,fung2017too}.

In addition to experiments, the response of molecular junctions to light has been extensively studied in theory. When the lead-molecule coupling is weak, the most efficient approach is to exploit quantum master equations\cite{may2008optical1,may2008optical2,muralidharan2006probing}. The effect of the resonant light is included into the master equation by considering a transition rate between ground and excited states, whose magnitude is proportional to the light intensity\cite{fu2018,fu2019photoinduced}. In most of theoretical studies the effect of triplet states are ignored due to the small triplet-singlet transition rate\cite{wang2011laser,may2008optical2,wang2010charge}. However, it was shown that the triplet states cannot be ignored in the studying the optical response of the molecular junctions, as even though the direct singlet-triplet transition is impossible, one can adjust the parameters in such a way to have the indirect transition through the charged states\cite{fu2018,fu2019photoinduced}.

In this work we consider the OPE-3 molecule bridging two metallic leads and show that by adjusting the gate voltage one can have the novel phenomena of light induced NDC. This effect is explained using the transitions between the charged and triplet states in the presence of the e-ph interactions and light. It should be noted that in addition to the value for fundamental physics, such behaviors can be very important for engineering future switches and transistors in molecular electronics. In our work however, we predict the additional property of being controllable by light. The rest of this document is organized as follow. In Sec.\ref{sec:mm} we describe our theoretical method and present the corresponding master equations as well as current formula. In Sec.\ref{sec:res} we consider the OPE-3 molecule and using \textit{ab initio} calculations obtain its phonon mode energies and their corresponding displacements due to electronic state transitions. In order to make the calculations affordable we map nearly degenerate modes into single effective modes, the theory of which is described in the Ref.\cite{eskandariasl2020role}. We show that by adjusting the gate voltage one can have a light induced NDC and describe the roles of triplet states, e-ph interaction and light to have this phenomena. Moreover, we investigate the effects of light intensity and temperature. Finally, Sec.\ref{sec:conc} concludes our work.

\section{model and method} \label{sec:mm}
\begin{figure} 
\includegraphics[width=\columnwidth]{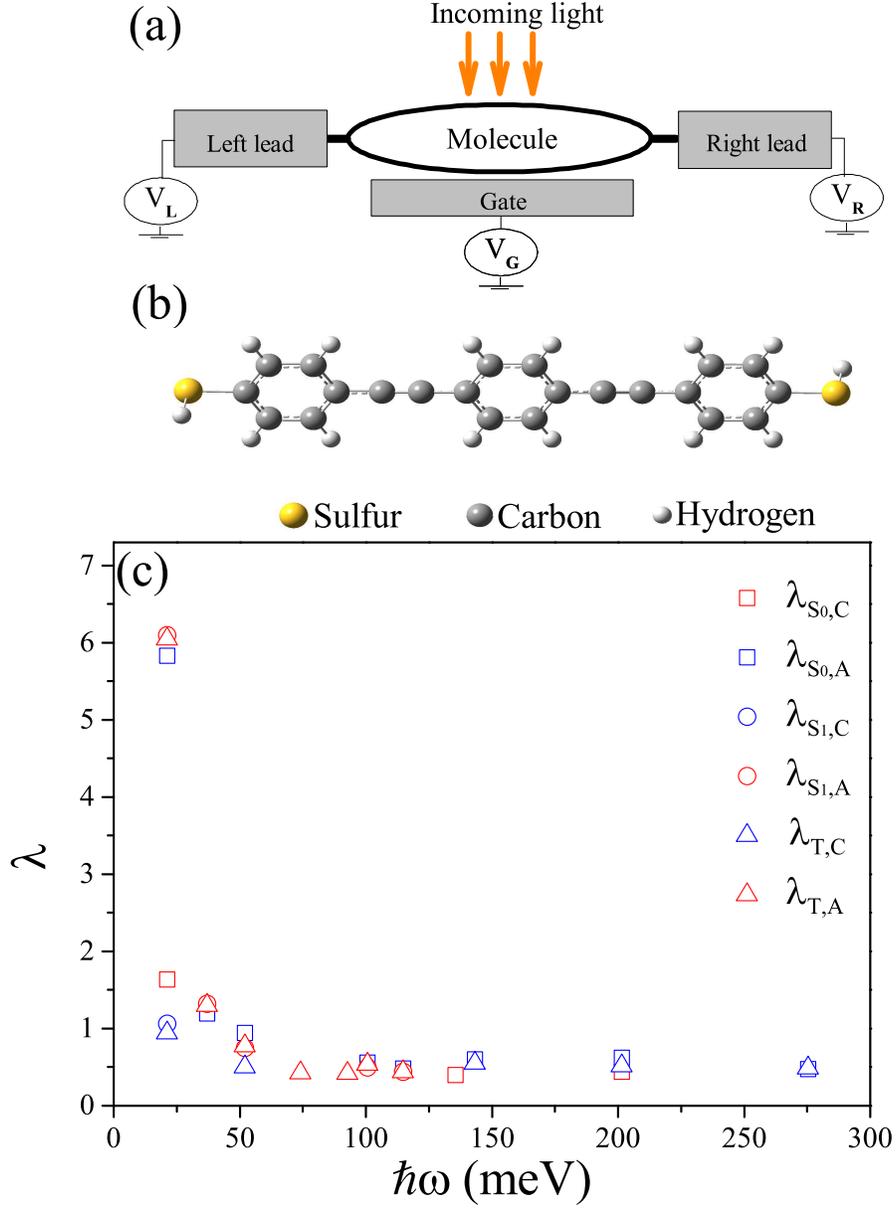}\caption{(Color online) (a) Schematic setup. The illuminated molecule is bridging the left and right metallic electrodes while capacitively coupled to the gate electrode. (b) The molecule OPE-3 with the anchoring sulfur atoms. (c) The results of first principle calculations for the electron-phonon coupling strengths. Phonon modes with the energy differences less than 3meV (equivalent to 35 K) are combined and electron-phonon couplings less than 0.4 are ignored.}
\label{smd}%
\end{figure}

Our setup is composed of a molecular junction bridging two leads and capacitively coupled to a gate electrode. The electrons on the molecule are coupled to its vibrational modes which in turn can be dissipated to a thermal bath. We also consider an incoming classical light beam which interacts with the electrons on our molecule. A schematic representation of this setup is shown in Fig.\ref{smd} a. The total Hamiltonian is 
\begin{eqnarray}
    \hat{H}=\hat{H}_{\mathrm{mol}}+\hat{H}_{\mathrm{env}}+\hat{H}_{\mathrm{coup}}+\hat{H}_{\mathrm{field}},
    \label{hamtot}
\end{eqnarray}
where $\hat{H}_{\mathrm{mol}}$ is for the molecule, $\hat{H}_{\mathrm{env}}$ describes the environment involving leads and phonon thermal bath, $\hat{H}_{\mathrm{coup}}$ is responsible for the copuling between the molecular junction and the environment, and $\hat{H}_{\mathrm{field}}$ is the interaction Hamiltonian between the incoming light and the molecule. 

The molecular Hamiltonian which involves both electrons and phonons and their interaction, is given by
\begin{eqnarray}
&\hat{H}_{\mathrm{mol}}=\sum_{M} \sum_{r=1}^{R} \vert M\rangle \langle M \vert  \nonumber\\
&\biggr(\xi_{M}+\hbar \omega_{r}\left[ \left( \hat{b}_{r}^{\dagger}-\lambda_{r,M}\right)\left( \hat{b}_{r}-\lambda_{r,M}\right)+1/2 \right]  \biggr),
\label{h-vib}
\end{eqnarray}
where $ \{\vert M \rangle \} $ is the set of many-electron states of the molecule. The electronic energy of the many-electron state $ \vert M\rangle $ is $ \xi_{M}=E_{M}-N_M e V_{G} $ with $E_{M}$ being the eigenenergy of the electronic Hamiltonian, $N_M$ the number of excess electrons in state $ \vert M\rangle $ with respect to the neutral state, and $ V_{G} $ the applied gate voltage. $ \hat{b}_{r}^{\dagger} $ ($ \hat{b}_{r} $) denotes the creation (annihilation) operator of the $r$th phonon mode, while the total number of phonon modes is $R$. Finally, $\lambda_{r,M}$ is the displacement of the $r$th phonon mode associated with the electronic state $\vert M\rangle$.

The environment Hamiltonian, $\hat{H}_{\mathrm{env}}=\hat{H}_{\mathrm{leads}}+\hat{H}_{\mathrm{th}}$, is the summation of two parts corresponding to electronic leads and phonon thermal bath, respectively. The leads are considered as non-interacting Fermi gases described by $ \hat{H}_{\mathrm{leads}}=\sum_{\alpha=L,R} \varepsilon_{\alpha k \sigma} \hat{c}_{\alpha k \sigma}^{\dagger} \hat{c}_{\alpha k \sigma} $, while the thermal bath is a continuum of quantum harmonic oscillators determined by $\hat{H}_{\mathrm{th}}=\sum_{q}  \hbar \omega_{q} \hat{b}_{q}^{\dagger} \hat{b}_{q}$. Here $ \hat{c}_{\alpha k \sigma}^{\dagger} $ ($ \hat{b}_{q}^{\dagger} $) creates an electron (phonon) with energy $ \varepsilon_{\alpha k \sigma} $ ($ \hbar \omega_{q} $) in the lead $ \alpha $ (thermal bath). The coupling Hamiltonian is $\hat{H}_{\mathrm{coup}}=\hat{H}_{\mathrm{m-l}}+\hat{H}_{\mathrm{m-th}}$, where the first term determines the hopping of electrons between the molecule and leads, $ \hat{H}_{\mathrm{m-l}}=\sum_{M,M^{\prime} ,\alpha k \sigma} V_{\alpha k \sigma ,M,M^{\prime}} c^{\dagger}_{\alpha k \sigma} \vert M^{\prime}\rangle \langle M \vert \delta_{N_{M},N_{M^{\prime}}+1} +h.c.$, while the second term indicates the coupling of the phonons to the thermal bath, $\hat{H}_{\mathrm{m-th}}=\sum_{q} \sum_{r} t_{q,r} (\hat{b}_{q}^{\dagger}+\hat{b}_{q}) (\hat{b}_{r}^{\dagger}+\hat{b}_{r})$. 

For our theoretical investigation we select the following many-electron states of the molecule. The ground and first excited singlet states which are shown, respectively, by $\vert S_0 \rangle$ and $\vert S_1 \rangle$, the triplet states which are denoted by $\vert T^m \rangle$ in which $m=0,\pm 1 $, and the charged anion, $\vert A^\sigma \rangle$, and cation,  $\vert C^\sigma \rangle$, states in which $\sigma=\pm \frac{1}{2} $. The singlet and triplet states are neutral, i.e., $N_{S_0}=N_{S_1}=N_{T^m}=0$, while for the charged states we have $N_{A^\sigma}=-N_{C^\sigma}=1$. The electron transport between the molecule and leads changes the molecule from the neutral state to the charged state and vice versa. The spin-orbit coupling could in principle cause transitions between singlet and triplet states with a very low rate which is ignored in our treatment. Finally, the incoming light can interchange the molecular state between $\vert S_0 \rangle$ and $\vert S_1 \rangle$, via the Hamiltonian\cite{schatz2002quantum}
\begin{eqnarray}
\hat{H}_{\mathrm{field}}=-\mathbf{E}(t).\mathbf{d}_{01} \vert S_0 \rangle \langle S_1 \vert + h.c.,     
\end{eqnarray}
where $\mathbf{E}$ is the electric field of the incoming light beam and $\mathbf{d}_{01}$ is the dipole moment between the ground and first excited singlet states. 

By exploiting a polaron transformation and considering Born-Markov approximation in accordance to the weak molecule-lead coupling (a straight forward generalization of the derivation given in Ref.\cite{fu2019photoinduced}) , the master equation in the Pauli form can be obtained as
\begin{eqnarray}
\frac{dP_{M \mathbf{n}}}{dt}=\sum_{M^{\prime} \mathbf{n^{\prime}}} \left( k_{M\mathbf{n},M^{\prime} \mathbf{n^{\prime}} } P_{M^{\prime} \mathbf{n^{\prime}}} - k_{ M^{\prime} \mathbf{n^{\prime}}, M\mathbf{n} } P_{M\mathbf{n}} \right) , \label{dpav} 
\end{eqnarray}
where $P_{M \mathbf{n}}$ is the probability that the state of molecule be $\vert M \mathbf{n} \rangle$, in which $ \mathbf{n} $ is a $R$-dimensional vector with integer elements $n_r$ determining the number of phonons in the mode $r$. The transition rate contains contributions from the leads, thermal bath and the electric field of light, and can be written as $ k_{M\mathbf{n},M^{\prime} \mathbf{n^{\prime}} }=\sum_{\alpha=L,R}  k^{\alpha}_{M\mathbf{n},M^{\prime} \mathbf{n^{\prime}} }+  k^{\rm{th}}_{M\mathbf{n},M^{\prime} \mathbf{n^{\prime}} }+k^{\rm{field}}_{M\mathbf{n},M^{\prime} \mathbf{n^{\prime}} }$. The first term, which is the transition induced by the lead $\alpha$, is given by
\begin{eqnarray}
 k^{\alpha}_{M\mathbf{n},M^{\prime} \mathbf{n^{\prime}} }=\Lambda^{\alpha}_{M\mathbf{n},M^{\prime} \mathbf{n^{\prime}} } \prod_{r=1}^{R} \chi_{n^{\prime}_{r} n_{r}} \left(\lambda_{r,MM^{\prime}}\right)   
 \label{kalnn}
\end{eqnarray}
where 
\begin{eqnarray}
&&\Lambda^{\alpha}_{M\mathbf{n},M^{\prime} \mathbf{n^{\prime}} }=\Gamma^{\alpha} \upsilon_{M, M^{\prime}} \biggr[  \delta_{N_M,N_{M^{\prime}}+1} f\left(\varepsilon_{M\mathbf{n}, M^{\prime} \mathbf{n^{\prime}}},\mu_{\alpha} \right) \nonumber\\
&&+\delta_{N_M,N_{M^{\prime}}-1} \left(1-f\left(\varepsilon_{M^{\prime} \mathbf{n^{\prime}} , M\mathbf{n}},\mu_{\alpha} \right)  \right) \biggr]
\label{lambv}
\end{eqnarray}
in which $ \Gamma^{\alpha} $ is the electron transfer rate between  molecular junction and the lead $ \alpha $, and $  \upsilon_{M, M^{\prime}}=\sum_{i} \vert \langle M \vert \hat{c}_i \vert M^{\prime} \rangle \vert^2+ \sum_{i} \vert \langle M \vert \hat{c}^{\dagger}_{i} \vert M^{\prime} \rangle \vert^2 $, where $ i $ runs over any complete single-particle basis (including spin) for electrons on the molecule. Moreover, $ \delta_{\mu,\nu} $ is the Kronecker delta and $f(\varepsilon,\mu)=1/(1+e^{(\varepsilon-\mu)\beta })$ is the Fermi distribution function with $ \beta=1/k_B T $ being the reciprocal temperature. $ \varepsilon_{M\mathbf{n}, M^{\prime} \mathbf{n^{\prime}}}= \xi_{M}+\epsilon_{\mathbf{n}}-\xi_{M^{\prime}}-\epsilon_{\mathbf{n^{\prime}}} $, in which $\epsilon_{\mathbf{n}}=\sum_{r=1}^{R} \hbar \omega_{r} n_{r}$ , and $ \mu_{\alpha}=\mu_0 + e V_{\alpha} $ is the chemical potential of the lead $ \alpha $ (the applied bias voltage is $ V=(\mu_L-\mu_R)/e=V_L-V_R $). Moreover, $\lambda_{r,MM^{\prime}}=\vert \lambda_{r,M}-\lambda_{r,M^{\prime}} \vert $, is the dimensionless e-ph coupling of the $ r $th phonon mode. Finally,    
\begin{eqnarray}
&&\chi_{n^{\prime} n} \left(\lambda \right) =\vert \langle n \vert e^{\lambda (\hat{b}^{\dagger}-\hat{b}) } \vert n^{\prime} \rangle \vert^{2}= \nonumber\\
&&e^{-\lambda^{2}} \biggr\vert\sum_{j=0}^{n_{<}} \frac{(-1)^{j}\lambda^{2j+n_{>}-n_{<}} \sqrt{n!n^{\prime}!}}{j!(j+n_{>}-n_{<})!(n_{<}-j)!}\biggr\vert^{2} ,
\label{xn}
\end{eqnarray}
in which $n_{>(<)}$ is the maximum (minimum) of $n$ and $n^{\prime}$.

The coupling to the thermal bath results in thermal excitation-dissipation of phonons with the corresponding rate given by
\begin{eqnarray}
&&  k^{th}_{M \mathbf{n},M^\prime \mathbf{n^\prime}}=\delta_{M,M^\prime}  \gamma^{th} \sum_{r=1}^{R} \biggr[ \delta_{n_r,n^{\prime}_r+1}(n^{\prime}_r+1) n_{th}(\omega_r)\nonumber \\ 
&&+\delta_{n_r,n^{\prime}_r-1} n^{\prime}_r  (n_{th}(\omega_r)+1) \biggr], 
\end{eqnarray}
where $ \gamma^{th} $ determines the relaxation rate of phonons to the thermal bath and $n_{th}(\omega_r)=1/(e^{\beta \hbar \omega_r}-1)$.

The incoming light can only cause a transition between the states $\vert S_0 \rangle$ and $ \langle S_1 \vert$ with the rate given by
\begin{eqnarray}
 k^{\rm{field}}_{S_0 \mathbf{n},S_1 \mathbf{n^{\prime}} }=k^{\rm{field}}_{S_1 \mathbf{n},S_0 \mathbf{n^{\prime}} } =k^{\rm{field}} \prod_{r=1}^{R} \chi_{n^{\prime}_{r} n_{r}} \left(\lambda_{r,MM^{\prime}}\right),   
\end{eqnarray}
where the instantaneous emission rate is neglected compared to $k^{\rm{field}}$, which is the optical induced transition rate tunable with the light intensity\cite{fu2019photoinduced}.  

In this work we consider the phonons to be strongly coupled to the thermal bath so that $\gamma^{th}$ is much larger than the transition rates induced by light and leads. In such a case phonon population is instantly equilibrated so that $P_{M,\mathbf{n}}=e^{-\beta \epsilon_{\mathbf{n}} } P_{M,\mathbf{0} }$. As the phonon distribution is now thermal, the total population of electronic states, $P_{M}=\sum_{\mathbf{n}} P_{M,\mathbf{n}} $ should be investigated rather than each individual $P_{M,\mathbf{n} }$. It's straightforward to show that the dynamics of these equilibrated probability distributions is given as $dP_M /dt=\sum_{M^{\prime} } \left( k_{M,M^{\prime} } P_{M^{\prime}} - k_{ M^{\prime} , M } P_{M} \right)$, where $  k_{M,M^{\prime} }=\sum_{\alpha=L,R}  k^{\alpha}_{M,M^{\prime}}+ k^{\rm{field}}_{M,M^{\prime} }$, where
\begin{eqnarray}
k^{\alpha}_{ M,M^{\prime} } = \frac{1}{Z} \sum_{\mathbf{n},\mathbf{n^{\prime}}} \Lambda^{\alpha}_{M\mathbf{n}, M^{\prime} \mathbf{n^{\prime}} } \prod_{r=1}^{R} \chi_{n^{\prime}_{r} n_{r}} \left(\lambda_{r,MM^{\prime}}\right)    e^{-\beta \epsilon_{\mathbf{n^{\prime}}}}
\label{keqv}
\end{eqnarray}
in which
\begin{eqnarray}
&&Z=\sum_{\mathbf{n}} e^{-\beta \epsilon_{\mathbf{n}}}=\prod_{r=1}^{R} \frac{1}{1-e^{-\beta \hbar \omega_{r}}},
\end{eqnarray}
and
\begin{eqnarray}
k^{\rm{field}}_{M,M^{\prime} }=k^{\rm{field}} \left( \delta_{M,S_0} \delta_{M^{\prime},S_1 } + \delta_{M,S_1} \delta_{M^{\prime},S_0 } \right) .  
\end{eqnarray}

Investigating the electron population change, one can show that the electrical current from lead $\alpha$ to the molecule is
\begin{eqnarray}
I^{\alpha}= e \sum_{M,M^{\prime} }  P_{M} k^{\alpha}_{M^{\prime}, M } \left( \delta_{N_M+1,N_{M^{\prime}}}  - \delta_{N_M-1,N_{M^{\prime}}}  \right). \quad
\label{ieq}
\end{eqnarray}

\section{Results and discussions} \label{sec:res}
We consider the OPE-3 molecule with the structure as shown in Fig.\ref{smd} b. Using the Gaussian 09 software at the B3LYP\cite{becke1993} level and exploiting the cc-pVDZ basis set, the geometry optimization and phonon mode energies are calculated. Moreover, performing the Duschinsky transformation\cite{duschinsky} using the DUSHIN\cite{DUSHIN} code, we computed the e-ph couplings for all possible charge transitions in the truncated Fock space of molecular electronic states. For a thermal bath at the temperature of 35 K, the phonon modes with the energy differences less than 3 meV can be considered as nearly-degenerate. For the transport calculations, nearly-degenerate modes can be mapped into an effective mode with the effective Huang-Rhys factor (which is the square of e-ph coupling strength) being the average of the Huang-Rhys factors of the original modes\cite{eskandariasl2020role}. The effective resulting e-ph couplings are shown in Fig. \ref{smd} c, where the effective couplings less than 0.4 are ignored.

The electronic state energies are obtained to be $E_{S_0}=0$, $E_{S_1}=3.029$ eV, $E_{T}=1.871$ eV, $E_{C}=6.178$ eV and $E_{A}=-1.365$ eV, in which for the anion and cation states an image charge correction of -0.3 eV is considered\cite{fu2019photoinduced}. Moreover, the non-biased chemical potential of the metallic leads is assumed to be $\mu_0=-5.3$ eV. The charge transport through molecule can be controlled by applying a gate voltage. The gate voltage that aligns triplet states with anion states is $V_{G}^{T,A}=(E_A-E_T-\mu_0)/e=2.064$ V (the concept of state alignment is described in the Refs. \cite{fu2018,fu2019photoinduced}) and in our investigations other gate voltages are measured with respect to $V_{G}^{T,A}$. In our calculations we consider a symmetric lead-molecule coupling with the rate $\Gamma=\Gamma^{R}=\Gamma^{L}=0.1$ meV, and the bias voltage is applied symmetrically, $V_L=-V_R=V/2$. 

\begin{figure} 
\includegraphics[width=\columnwidth]{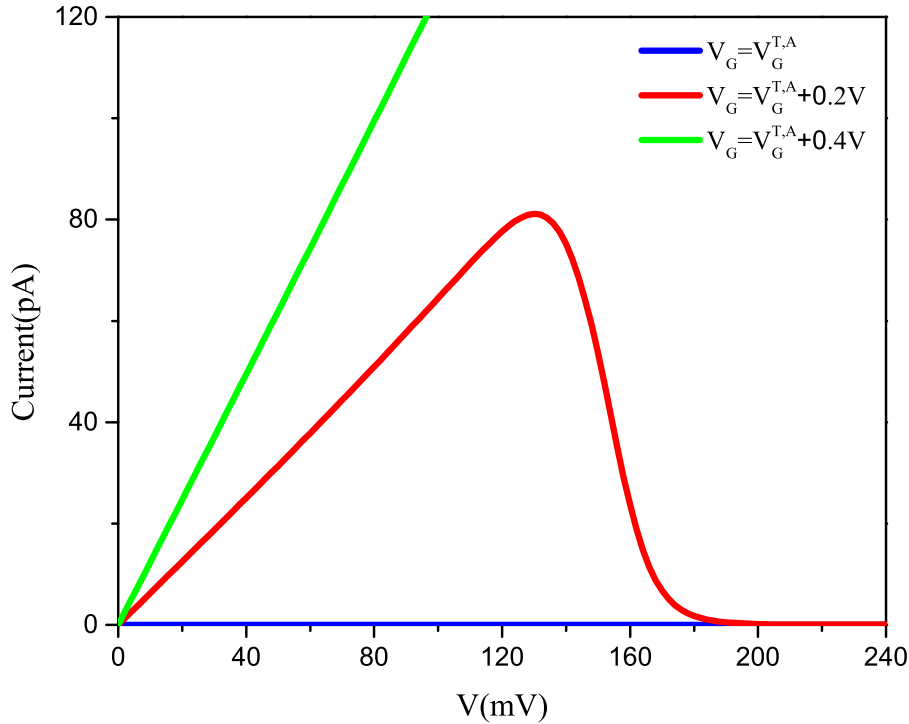}\caption{(Color online) IV curves for several values of the gate voltages. In order to have NDC we have to adjust the gate voltage to be slightly more than $V_{G}^{T,A}$, as in the red curve here. The light induced transition rate is $k^{\rm{field}}=10^{12} ~ s^{-1} $  and the temperature is 35 K, which corresponds to $k_B T=3$ meV. }
\label{vgs}%
\end{figure}

In Fig. \ref{vgs} we considered three values of the gate voltage and obtained the corresponding IV curves. The light induced transition rate is $k^{\rm{field}}=10^{12} ~ s^{-1} $  and the temperature is 35 K, which corresponds to $k_B T=3$ meV. It's seen that for the gate voltage $V_{G}=V_{G}^{T,A}$ we have no current at all. In order to understand this behavior, we have to investigate the state transition rates. At this gate voltage the important transitions are $S_0 \leftrightarrows S_1 \rightarrow C \rightarrow T  $, as a result of which the molecule is totally populated with the triplet states and no steady-state current can be established. In order to have an electrical current in this case, one has to increase the bias voltage to overcome the Franck-Condon blockade and make the transition $T \rightarrow A$ possible (not shown in the figure). 

By increasing the gate voltage, as it is the case for $V_{G}=V_{G}^{T,A}+0.2$ V and $V_{G}=V_{G}^{T,A}+0.4$ V in Fig. \ref{vgs}, the transition $S_1 \rightarrow C$ is blocked for low bias voltages and the cation state population vanishes, which in turn makes the triplet states unreachable, as only the cation states can make transition to them in this regime. Consequently, it is possible to have a low bias current of the order of 10-100 pA which is measurable in experimental setups\cite{kubatkin2003}. The steady-state current is now maintained by the transitions $S_0 \leftrightarrows S_1 \rightarrow A \rightarrow S_0  $. One should notice that because of the Franck-Condon blockade resulting from e-ph coupling, the transition rates in and out of the anion states are different depending on the lead which induces them. In short,  $K^L_{A,S_1} \neq K^R_{A,S_1}$ and $K^L_{S_0,A} \neq K^R_{S_0,A}$, which is the reason of having current and determines the important role of phonons.  

For the gate voltage $V_{G}=V_{G}^{T,A}+0.2$ V ( the red curve in Fig. \ref{vgs}), we have NDC as the current is decreased by increasing the bias voltage above $V \simeq 130 $mV. This behavior is explained by noticing that at higher bias voltages a transition from the state $S_1$ to the cation states gradually turns on, which in turn decay to the triplet states and make them populated. Because of the Franck-Condon blockade, the triplet states make no transitions to the charged states which means electrons cannot jump into or outside of the molecule. Consequently, the current is blocked and NDC emerges. It is clear from this discussion that in order to have NDC we need both e-ph coupling and the presence of triplet states. Moreover, it is noteworthy that this phenomenon occurs at low bias voltages, hence the phonon modes with low energies play the most important role. 

\begin{figure} 
\includegraphics[width=\columnwidth]{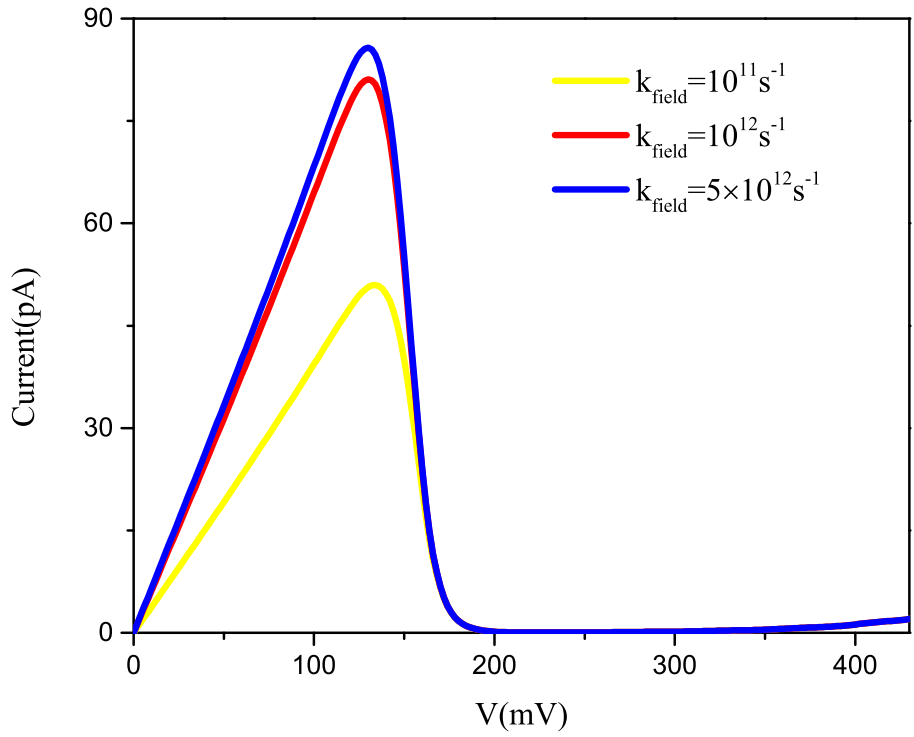}\caption{(Color online) IV curves for several values of the light induced transition rates. Reducing the light intensity reduces $k^{\rm{field}}$ which lowers the peak in IV curve which means weakening the NDC. The gate voltage is $V_{G}^{T,A}+0.2$ V and the temperature is 35 K. }
\label{ks}%
\end{figure}

In addition to e-ph coupling, the interaction with the incoming classical light is crucial for having NDC. This is explicitly shown in Fig.\ref{ks}, where at the fixed gate voltage of $V_{G}=V_{G}^{T,A}+0.2$ and temperature of 35K, we considered several values of the light induced transitions, $k^{\rm{field}}$. Reducing the light intensity reduces the transition rates of the $S_0 \leftrightarrows S_1$ processes, therefore, the probability that the molecule stays in the neutral state $S_0$ and maintains no electrical current increases. On the other hand, by increasing the light intensity the peak value of current reaches its saturated value, as other transitions (i.e., $S_1 \rightarrow A \rightarrow S_0  $) are not affected. Moreover, according to our former discussions the bias voltage at which the NDC starts is not determined by the $S_0 \leftrightarrows S_1$ processes which makes it independent of the light intensity. 

\begin{figure} 
\includegraphics[width=\columnwidth]{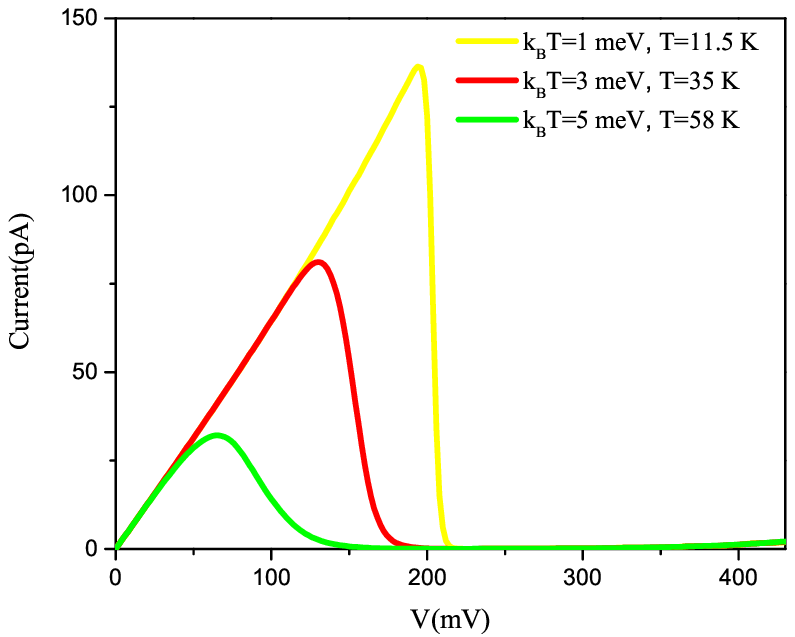}\caption{(Color online) IV curves for several values of temperatures. Lowering the temperature strengthens the NDC, while it still persists by increasing temperature. The gate voltage is $V_{G}^{T,A}+0.2$V and the light induced transition rate is $k^{\rm{field}}=10^{12} ~ s^{-1} $. }
\label{ts}%
\end{figure}

In order to investigate the effect of temperature in NDC, in Fig. \ref{ts} we show the IV curves for three different temperatures with the fixed gate voltage of $V_{G}=V_{G}^{T,A}+0.2$V and the light induced transition rate of $k^{\rm{field}}=10^{12} ~ s^{-1} $. It should be noted that our results for the case T=11.5 K are rough, since we have done the effective combination for the temperatures more than or equal to 35 K. Our results indicate that by decreasing the temperature the bias voltage at which we have NDC is increased, as well as the peak value of current. As we already explained, the NDC starts at the bias voltage where the $S_1 \rightarrow C$ transition becomes possible. At zero temperature the Fermi distribution function is a step function and this bias voltage (which is the peak voltage in the IV curve) can be obtained by demanding $K^R_{C,S_1}=0$ which results in $V=2(E_C - E_{S_1} + e V_G + \mu_0)/e=226 $mV. By increasing the temperature the step in the Fermi function gets smooth, which makes the $S_1 \rightarrow C$ transition possible at lower bias voltages and decreases the peak value of current.

\section{Conclusions} \label{sec:conc}
In conclusion, we studied the OPE-3 molecular junction and showed that the transition to triplet states, e-ph interaction and light illumination can result in the novel phenomena of light induced NDC. By adjusting the gate voltage, one can see that there is a bias voltage after which a transition from the excited singlet states to the cation states is possible, which in turn decay to the triplet states and block the light induced electronic current. Consequently, we have NDC. We showed that by reducing the light intensity the current peak is decreased while the bias voltage at which NDC starts is fixed. On the other hand, increasing the light intensity we approach the saturated current and NDC values. By increasing the temperature the step functions of the Fermi distribution functions of the leads get smooth, which make the transition to the cation state and consequently the NDC possible at lower temperatures. Our results can be of potential importance for constructing future controllable molecular switches and transistors.  

\section*{ACKNOWLEDGMENT}
We acknowledge useful discussions with Liang-Yan Hsu, Bo Fu, and Qian-Rui Huang.

\bibliographystyle{model1-num-names}
\bibliography{mp-017.bib}

\end{document}